\documentstyle[prl,aps,twocolumn]{revtex}

\begin{document}

\preprint{hep-ph/0004051}
\pagestyle{headings}

\title{Primordial Galactic Magnetic Fields from Domain Walls at
  the QCD Phase Transition.}

\author{Michael McNeil Forbes and  Ariel Zhitnitsky}
\address{
  Department of Physics and Astronomy, University of British Columbia\\
  Vancouver, British Columbia, Canada V6T 1Z1.}

\twocolumn[\hsize\textwidth\columnwidth\hsize
\csname@twocolumnfalse\endcsname
\maketitle

\begin{abstract}
  We propose a mechanism to generate large-scale magnetic fields with
  correlation lengths of $100$ kpc.  Domain walls with QCD-scale
  internal structure form and coalesce obtaining Hubble-scale
  correlations while aligning nucleon spins.  Because of strong CP
  violation, the walls are ferromagnetic, which induces
  electromagnetic fields with Hubble size correlations.  The CP
  violation also induces a maximal helicity (Chern-Simons) which
  supports an inverse cascade, allowing the initial correlations
  to grow to $100$ kpc today.  We estimate the generated
  electromagnetic fields in terms of the QCD parameters and discuss
  the effects of the resulting fields.
\end{abstract}
] 
\narrowtext
\paragraph*{Introduction.}
The source of cosmic magnetic fields with large scale correlations has
remained somewhat of a mystery \cite{observations}.  There are two
possible origins for these fields: primordial sources and galactic
sources.  Primordial fields are produced in the earlier universe and then
evolve and are thought to provide seeds which gravitational dynamos
later amplify.  Galactic sources would produce the fields as well as
amplify them.  Many mechanisms have been proposed
\cite{magreview,Cornwall:1997,Son:1999,FC:1998}, however, most fail to
convincingly generate fields with large enough correlation lengths to
match the observed microgauss fields with $\sim 100$ kpc correlations.
We present here a mechanism which, although probably requiring a
dynamo to produce microgauss fields, generates fields with hundred
kiloparsec correlations.  We present this mechanism as an application
of our recent understanding of QCD domain walls, which will be
described in detail elsewhere \cite{FZ:2000}.

\begin{enumerate}
  \setlength{\parskip}{-3pt}
\item Sometime near the QCD phase transition, $T_{\mathrm{QCD}}\approx
  1$ GeV, QCD domain walls form.
\item These domain walls rapidly coalesce until there remains, on
  average, one domain wall per Hubble volume with Hubble-scale
  correlations.
\item Baryons interact with the domain walls and align their spins
  along the domain walls.
\item The magnetic and electric dipole moments of the baryons induce
  helical magnetic fields correlated with the domain wall.
\item The domain walls decay, leaving a magnetic field.
\item As the universe expands, an ``inverse cascade'' mechanism
  transfers energy from small to large scale modes, effectively
  increasing the resulting correlation length of the observed large
  scale fields.
\end{enumerate}
We shall start by discussing the ``inverse cascade'' mechanism which
seems to be the most efficient mechanism for increasing the
correlation length of magnetic turbulence.  After presenting some
estimates to show that this mechanism can indeed generate fields of
the observed scales, we shall discuss the domain wall mechanism for
generating the initial fields.

\paragraph*{Evolution of Magnetic Fields.}
As suggested by Cornwall \cite{Cornwall:1997}, discussed by Son
\cite{Son:1999} and confirmed by Field and Carroll \cite{FC:1998},
energy in magnetic fields can undergo an apparent ``inverse cascade''
and be transfered from high frequency modes to low frequency modes,
thus increasing the overall correlation length of the field faster
than the na\"\i{}ve scaling by the universe's scale parameter $R(T)$.
There are two important conditions: turbulence must be supported as
indicated by a large Reynolds number ${\rm Re}$, and magnetic helicity
(Abelian Chern-Simons number) $H=\int{\vec{\bf A}\cdot\vec{\bf B}}{\rm
  d}^3{x}$ is approximately conserved.  The importance of helicity was
originally demonstrated by Pouquet and collaborators \cite{Pouquet}.
The mechanism is thus: the small scale modes dissipate, but the
conservation of helicity requires that the helicity be transfered to
larger scale modes. Some energy is transfered along with the helicity
and hence energy is transported from the small to large scale modes.
This is the inverse cascade.  The reader is referred to
\cite{Cornwall:1997,Son:1999,FC:1998} for a more complete discussion.

In the early universe, Re is very large and supports
turbulence.  This drops to $\mathrm{Re}\approx 1$ at the $e^+e^-$
annihilation epoch, $T_0\approx 100$ eV \cite{Son:1999}.  After this
point (and throughout the matter dominated phase) we assume that the
fields are ``frozen in'' and that the correlation length expands as
$R$ while the field strength decays as $R^{-2}$.  Note that the
inverse cascade is only supported during the radiation dominated
phase of the universe.

Under the assumption that the field is maximally helical, these
conditions imply the following relationships between the initial field
$B_{\mathrm{rms}}(T_i)$ with initial correlation $l(T_i)$ and present
fields today ($T_{\mathrm{now}}\approx2\times10^{-4}$ eV)
$B_{\mathrm{rms}}(T_{\mathrm{now}})$ with correlation
$l(T_{\mathrm{now}})$ \cite{Son:1999,FC:1998}:
\begin{eqnarray}
  B_{\mathrm{rms}}(T_{\mathrm{now}})&=&
  \left(\frac{T_0}{T_{\mathrm{now}}}\right)^{-2}
  \left(\frac{T_i}{T_0}\right)^{-7/3}B_{\mathrm{rms}}(T_i)
  \label{eq:brms1}\\
  l(T_{\mathrm{now}})&=&\left(\frac{T_0}{T_{\mathrm{now}}}\right)
  \left(\frac{T_i}{T_0}\right)^{5/3}l(T_i).
  \label{eq:corr1}
\end{eqnarray}

As pointed out in \cite{Son:1999}, the only way to generate turbulence
is either by a phase transition $T_i$ or by gravitational
instabilities.  We consider the former source.  As we shall show, our
mechanism generates Hubble size correlations $l_i$ at a phase
transition $T_i$.  In the radiation dominated epoch, the Hubble size
scales as $T_i^{-2}$.  Combining this with (\ref{eq:corr1}), we see
that $l_{\mathrm{now}}\propto T_i^{-1/3}$; thus, the earlier the phase
transition, the smaller the possible correlations.

The last phase transition is the QCD transition,
$T_i=T_{\mathrm{QCD}}\approx 0.2$ GeV with Hubble size
$l(T_{\mathrm{QCD}})\approx 30$ km.  We calculate (\ref{eq:EB}) the
initial magnetic field strength to be $B_{\mathrm{rms}}(T_i)\approx
e\Lambda_{QCD}^2/(\xi\Lambda_{\rm QCD})\approx (10^{17}{\rm
  G})/(\xi\Lambda_{\rm QCD})$ where $\xi$ is a correlation length that
depends on the dynamics of the system as discussed below and
$\Lambda_{\rm QCD}\approx 0.2$ GeV.  With these estimates, we see that
\begin{equation}
  \label{*}
  B_{\mathrm{rms}}\sim
  \frac{10^{-9}{\rm  G}}{\xi\Lambda_{\rm QCD}} ,~~~~ l\sim 100~{\rm kpc}
\end{equation}
today.  One could consider the electroweak transition which might
produce $100$ pc correlations today, but this presupposes a mechanism
for generating fields with Hubble-scale correlations.  Such a
mechanism does not appear to be possible in the Standard Model.
Instead, the fields produced are correlated at the scale $T_i^{-1}$
which can produce only $\sim 1$ km correlations today.

These are crude estimates, and galactic dynamos likely amplify these
fields.  The important point is that we can generate easily the $100$
kpc correlations observed today {\em provided} that the fields were
initially of Hubble size correlation.  Unless another mechanism for
amplifying the correlations of magnetic fields is discovered, we
suggest that, in order to obtain microgauss fields with $100$ kpc
correlation lengths, helical fields must be generated with
Hubble-scale correlations near or slightly after the QCD phase
transition $T_{\mathrm{QCD}}$.  The same conclusion regarding the
relevance of the QCD scale for this problem was also reached in
\cite{Son:1999,FC:1998}.  The rest of this work presents a mechanism
that can provide the desired Hubble size fields, justifying the
estimate (\ref{*}).  We shall explain the mechanism and give simple
estimates here.  See \cite{FZ:2000} for details.

\paragraph*{Magnetic field generation mechanism.}
The key players in our mechanism are domain walls formed at the QCD
phase transition that possess an internal structure of QCD scale.  We
shall present a full exposition of these walls in \cite{FZ:2000} but
to be concrete, we shall discuss an axion wall similar to that
described by Huang and Sikivie \cite{HS:1985}.

We start with a similar effective Lagrangian to that used by Huang and
Sikivie except we included the effects of the $\eta'$ singlet field
which they neglected:
\begin{equation}
  \label{eq:leff}
  {\cal L}_{{\rm eff}}=
   \frac{f_a^2}{2}\left|\partial_\mu e^{i\tilde{a}}\right|^2+
   \frac{f_{\pi}^2}{4}{\rm Tr}\left|\partial_\mu{\bf U}\right|^2-
   V({\bf U},\tilde{a})
\end{equation}
where $\tilde{a}=f_a^{-1}a$ is the dimensionless axion field and the
matrix ${\bf U}=\exp(i\tilde{\eta}'+i\tilde{\pi}^f{\bf \lambda}^f)$
contains the pion and $\eta'$ fields (to simplify the calculations, we
consider only the $SU(2)$ flavor group).  Although the $\eta'$ field is not
light, it couples to the anomaly and is the dominant player in
aligning the magnetic fields.  The potential
\begin{equation}
  \label{potential}
  V =\frac{1}{2} {\rm Tr}\left({\bf MU}e^{ i\tilde{a }}+{\rm h.c.}\right)-
  E\cos\left(\frac{i\ln(\det({\bf U}))}{
      N_c}\right)
\end{equation}
was first introduced in \cite{HZ}.  It should be realized that
$i\ln(\det({\bf U}))\equiv i\ln(\det({\bf U}))+2\pi n$ is a
multivalued function and we must choose the minimum valued branch.
Details about this potential are discussed in \cite{FZ:2000,HZ} but
several points will be made here.  All dimensionful parameters are
expressed in terms of the QCD chiral and gluon vacuum condensates, and
are well known numerically: ${\bf M} = -{\rm diag} (m_{q}^{i} |\langle
\bar{q}^{i}q^{i} \rangle| )$ and $ E = \langle b \alpha_s /(32 \pi)
G^2 \rangle$.  This potential correctly reproduces the
Veneziano-Witten effective chiral Lagrangian in the large $N_c$ limit
\cite{Wit2}; it reproduces the anomalous conformal and chiral Ward
identities of QCD; and it reproduces the known dependence in $\theta$
for small angles \cite{Wit2}.  We should also remark that the
qualitative results do not depend on the exact form of the potential:
domain walls form naturally because of the discrete nature of the
symmetries \cite{HS:1985,CHS:1999,dw}.

The result is that two different types of axion domain walls form
\cite{FZ:2000}.  One is almost identical to the one discussed in
\cite{HS:1985} with small corrections due to the $\eta'$.  We shall
call this the axion/pion ($a_{\pi}$) domain wall.  The second type,
which we shall call the axion/eta' ($a_{\eta'}$) domain wall is a new
solution characterized by a transition in both the axion and $\eta'$
fields.  The boundary conditions (vacuum states) for this wall are
$\tilde{a}(-\infty)=\tilde{\eta}'(-\infty)=0$ and
$\tilde{a}(\infty)=\tilde{\eta}'(\infty)=\pm\pi$ with $\pi^0=0$ at
both boundaries.  The main difference between the structures of the
two walls is that, whereas the $a_{\pi}$ domain wall has structure
only on the huge scale of $m_a^{-1}$, the $\eta'$ transition in
$a_{\eta'}$ has a scale of $m_{\eta'}^{-1}\sim\Lambda_{\rm QCD}^{-1}$.
The reason is that, in the presence of the non-zero axion ($\theta$)
field, the pion becomes effectively massless due to its Goldstone
nature.  The $\eta'$ is not sensitive to $\theta$ and so its mass
never becomes zero.  It is crucial that the walls have a structure of
scale $\Lambda_{\rm QCD}^{-1}$: there is no way for the $a_{\pi}$
wall to trap nucleons because of the huge difference in scales but the
$a_{\eta'}$ wall has exactly this structure and can therefore
efficiently align the nucleons.

The model we propose is this: Immediately after the phase transition,
the universe is filled with domain walls of scale
$T_{\mathrm{QCD}}^{-1}$.  As the temperature drops, these domain walls
coalesce, resulting in an average of one domain wall per Hubble
volume with Hubble-scale correlations \cite{CHS:1999,strings}.  It is
these $a_{\eta'}$ domain walls which align the dipole
moments of the nucleons producing the seed fields.

The following steps are crucial for this phenomenon: 1) The coalescing
of QCD domain wall gives the fields $\pi^f$, $\eta'$ Hubble-scale
correlations.  2) These fields interact with the nucleons producing
Hubble-scale correlations of nucleon spins residing in the vicinity of
the domain wall.  (The spins align perpendicular to the wall surface.)
3) Finally, the nucleons, which carry electric and magnetic moments
(due to strong CP violation), induce Hubble-scale correlated magnetic
and electric fields.  4) These magnetic and electric fields eventually
induce a nonzero helicity which has the same correlation.  This
helicity enables the inverse cascade.

\paragraph*{Quantitative Estimates.}
As outlined below, we have estimated the strengths of the induced
fields in terms of the QCD parameters \cite{FZ:2000}.  We consider two
types of interactions.  First, the nucleons align with the domain
wall.  Here we assume that the fluctuations in the nucleon field
$\Psi$ are rapid and that these effects cancel leaving the classical
domain wall background unaltered.  Thus, we are able to estimate many
mean values correlated on a large scale on the domain walls such as
$\langle\bar{\Psi}\gamma_5\sigma_{xy}\Psi\rangle$ and
$\langle\bar{\Psi}\gamma_z\gamma_5\Psi\rangle$ through the interaction
$\bar{\Psi}(i{\not{\!\partial}}-m_Ne^{i\tilde{\eta}'(z)\gamma_5})\Psi$.

To estimate the magnetization of the domain wall, we make the
approximation that the wall is flat compared to the lengths scales of
the nucleon interactions.  By assuming that momentum is conserved in
the wall, we reduce our problem to an effective $1+1$ dimensional theory
(in $z$ and $t$) which allows us to compute easily various mean value
using a bosonization trick \cite{JR,GW}.  The result for the mean
value $\langle\bar{\Psi}\gamma_5\sigma_{xy}\Psi\rangle$ for example is
\cite{FZ:2000}:
\begin{equation}
  \label{em3}
  \langle\bar{\Psi} \sigma_{xy}\gamma_{5}\Psi\rangle \simeq
  {\frac{\mu}{\pi}}\Lambda_{\rm QCD}^2,
\end{equation}
where $\mu\simeq m_N$ is a dimensional parameter originating from the
bosonization procedure of the corresponding 2D system and the
parameter $\Lambda_{\rm QCD}^2\sim\int {\rm d}k_x{\rm d}k_y$ comes
from counting the nucleon degeneracy in the $x$--$y$ plane of a Fermi
gas at temperature $T_c\simeq \Lambda_{\rm QCD}$.  These mean values
are only nonzero within a distance $\Lambda_{\rm QCD}^{-1}$ of the
domain wall and are correlated on the same Hubble-scale as the domain
wall.

From now on we treat the expectation value (\ref{em3}) as a background
classical field correlated on the Hubble-scale.  Once these sources
are known, one could calculate the generated electromagnetic field by
solving Maxwell's equations with the interaction
\begin{equation}
  \label{em4}
  {\cal L}_{\rm int}={\frac{1}{2}}(d_{\Psi}\bar{\Psi}
  \sigma_{\mu\nu}\gamma_5\Psi +\mu_{\Psi}\bar{\Psi}i \sigma_{\mu\nu}
  \Psi) F_{\mu\nu} + \bar{\Psi}(i{\rm D})^2\Psi
\end{equation}
where $d_{\Psi}$ ($\mu_{\Psi}$) is effective electric (magnetic)
dipole moments of the field $\Psi$.  Due to the CP violation (nonzero
$\theta$) along the axion domain wall, the anomalous nucleon dipole
moment in (\ref{em4}) $d_{\Psi}\sim \mu_{\Psi}\sim \frac{e}{m_N}$ is
also nonzero \cite{CDVW:1979}.  This is an important point: if no
anomalous moments were induced, then only charged particles could
generate the magnetic field: the walls would be diamagnetic not
ferromagnetic as argued in \cite{Voloshin2} and Landau levels would
exactly cancel the field generated by the dipoles.

Solving the complete set of Maxwell's equations, however, is extremely
difficult.  Instead, we use simple dimensional arguments.  For a small
planar region of area $\xi^2$ filled with aligned dipoles with
constant density, we know that the net magnetic field is proportional
to $\xi^{-1}$ since the dipole fields tend to cancel, thus for a flat
section of our domain wall, the field would be suppressed by a factor
of $(\xi\Lambda_{\rm QCD})^{-1}$.  For a perfectly flat, infinite
domain wall ($\xi\rightarrow \infty$), there would be no net field as
pointed out in \cite{Voloshin2}.  However, our domain walls are far
from flat.  Indeed, they have many wiggles and high frequency modes,
thus, the size of the flat regions where the fields are suppressed is
governed by a correlation $\xi$ which describes the curvature of the
wall.  Thus, the average electric and magnetic fields produced by the
domain wall are of the order
\begin{equation}
  \label{em5}
  \langle F_{\mu\nu}\rangle \simeq
  \frac{1}{\xi\Lambda_{\rm QCD}}\left[d_{\Psi}\langle \bar{\Psi}
    \sigma_{\mu\nu}\gamma_{5}\Psi\rangle +\mu_{\Psi}\langle \bar{\Psi}i
    \sigma_{\mu\nu} \Psi\rangle\right]
\end{equation}
where $\xi$ is an effective correlation length related to the size
of the dominant high frequency modes.

To estimate what effective scale $\xi$ has, however, requires an
understanding of the dynamics of the domain walls.  Initially, the
domain walls are correlated with a scale of $\Lambda_{\rm QCD}^{-1}$.
As the temperature cools, the walls smooth out and the lower bound
$\xi_-(t)$ for the scale of the walls correlations increases from
$\xi_-(0)\simeq\Lambda_{\rm QCD}^{-1}$.  This increase is a dynamical
feature, however, and is thus slow.  In addition, the walls coalesce
and become correlated on the Hubble-scale generating large scale
correlations.  Thus the wall has correlations from $\xi_-(t)$ up to the
upper limit set by the Hubble-scale.  We expect that
$\xi\ll$~Hubble~size at the time that the fields are aligned and that
the suppression is not nearly as great as implied in \cite{Voloshin2}.
Note that, even though the effects are confined to the region close to
the wall, the domain walls are moving and twisted so that the effects
occur throughout the entire Hubble volume.

The picture is thus that fields of strength
\begin{equation}
  \label{eq:EB}
  \langle E_z\rangle\simeq\langle B_z\rangle\sim 
  \frac{1}{\xi\Lambda_{\rm QCD}}\frac{e}{m_N}\frac{m_N\Lambda_{\rm QCD}^2}{\pi}
  \sim
  \frac{e\Lambda_{\rm QCD}}{\xi\pi}
\end{equation}
are generated with short correlations $\xi$, but then domains are
correlated on a large scale by the Hubble-scale modes of the
coalescing domain walls.  Thus, strong turbulence is generated with
correlations that run from $\Lambda_{\rm QCD}$ up to the Hubble-scale.

Finally, we note that this turbulence should be highly helical.  This
helicity arises from the fact that both electric and magnetic fields
are correlated together along the entire domain wall, $\langle\vec{\bf
  E}\rangle\sim\langle \vec{\bf A}\rangle / \tau$ where $\langle
\vec{\bf A}\rangle$ is the vector potential and $\tau$ is a relevant
timescale for the electric field to be screened (we expect
$\tau\sim\Lambda_{\rm QCD}^{-1}$ as we discuss below).  The magnetic
helicity density is thus
\begin{equation}
  \label{helicity}
  h\sim{\vec{\bf A}\cdot\vec{\bf B}} \sim\tau\langle E_z \rangle\langle
  B_z \rangle \sim \tau\frac{e^2}{\pi^2}\frac{\Lambda_{QCD}^2}{{\xi}^2}.
\end{equation}
Note carefully what happens here: The total helicity was zero in the
quark-gluon-plasma phase and remains zero in the whole universe, but
the helicity is separated so that in one Hubble volume, the helicity
has the same sign.  The reason for this is that, as the domain walls
coalesce, initial perturbations cause either a soliton or an
antisoliton to dominate and fill one Hubble volume.  In the
neighboring volume, there will be other solitons and antisolitons so
that there is an equal number of both, but they are spatially
separated which prevents them from annihilating.  This is similar to
how a particle and antiparticle may be created and then separated so
they do not annihilate.  In any case, the helicity is a pseudoscalar
and thus has the same sign along the domain wall: The entire Hubble
volume has helicity of the same sign.  This is the origin of the
Hubble-scale correlations in the helicity and in $B^2$.  The
correlation parameter $\xi$ which affects the magnitude of the fields
plays no role in disturbing this correlation.

Eventually, the electric field will be screened.  The time scale for
this is set by the plasma frequency for the electrons (protons will
screen much more slowly) $\omega_p\sim\Lambda_{\rm QCD}$.  The nucleons, however, also align on a similar
timescale $\Lambda_{QCD}^{-1}$, and the helicity is generated on this
scale too, so the electric screening will not qualitatively affect the
mechanism.  Finally, we note that the turbulence requires a seed which
remains in a local region for a timescale set by the conductivity
$\sigma\sim cT/e^2\sim\Lambda_{\rm QCD}$ where for $T=100$ MeV,
$c\approx 0.07$ \cite{Dimopoulos:1997nq} and is smaller for higher
$T$.  Thus, even if the domain walls move at the speed of light (due
to vibrations), there is still time to generate turbulence.

For this mechanism to work and not violate current observations, it
seems that the domain walls must eventually decay.  Several mechanisms
have been discussed for the decay of axion domain walls
\cite{CHS:1999,axion-review} and the timescales for these decays are
much larger than $\Lambda_{\rm QCD}^{-1}$, ie. long enough to generate
these fields but short enough to avoid cosmological problems.  QCD
domain walls \cite{FZ:2000} are quasistable and may nicely solve this
problem.  We assume that some mechanism exists to resolve the domain
wall problem in an appropriate timescale.  Thus, all the relevant
timescales are of the order $\Lambda_{\rm QCD}^{-1}$ except for the
lifetime of the walls and thus, although the discussed interactions
will affect the quantitative results, they will not affect the
mechanism or substantially change the order of the effects.

\paragraph*{Conclusion.}
We have shown that this mechanism can generate the magnetic fields
(\ref{*}) with large correlations, though galactic dynamos should
still play an important amplification role.  It seems that the crucial
conditions for the dynamo to take place are fields $B>10^{-20}$ G with
large ($100$ kpc) correlations.  From (\ref{*}) we see that we have a
huge interval $1\leq\xi\Lambda_{\rm QCD}\ll10^{10}$ of $\xi$ to seed
these dynamos.  Also, if $\xi$ is small, then this mechanism may
generate measurable extra-galactic fields.

We mention two new points that distinguish this mechanism from
previous proposals \cite{Iwazaki}.  First, the key nucleon is the
neutron which generates the fields due to an anomalous dipole moment
induced by the CP violating domain walls.  The nucleons thus make the
wall ferromagnetic, not diamagnetic as discussed in \cite{Voloshin2}.
Second, the interaction between the domain walls and nucleons are
substantial because of the QCD scale of the $\eta'$ transition.  There
is no way that axion domain walls with scales $\sim m_{a}^{-1}$ can
efficiently align nucleons at a temperature $T_{\rm QCD}$.

We should also note that the magnitudes of the fields generated by
this mechanism are small enough to satisfy the constraints placed by
nucleosynthesis and CMB distortions.  Thus, domain walls at the QCD
phase transition, in particular those described in \cite{FZ:2000},
provide a nice method of generating magnetic fields on $100$ kpc
correlations today (\ref{*}).
 
This work was supported by the NSERC of Canada.  We would like to
thank R. Brandenberger for many useful discussions.  AZ wishes to
thank: M. Shaposhnikov and I. Tkachev for valuable discussions which
motivated this study; Larry McLerran and D. Son for discussions on
Silk damping; and M. Voloshin and A. Vainshtein for discussions on the
magnetic properties of domain walls.

\bibliographystyle{prsty}

\end{document}